\documentstyle[12pt]{article}
\begin{document}
\begin{titlepage}
\begin{centering}

\vspace{2cm}

{\Large\bf Laboratory Constraints on a 33.9 MeV/c$^2$}\\
\vspace{0.5cm}
{\Large\bf Isosinglet Neutrino: Status and Perspectives}

\vspace{3cm}

Jan Govaerts\footnote{E-mail address: {\tt govaerts@fynu.ucl.ac.be}},
Jules Deutsch\footnote{E-mail address: {\tt deutsch@fynu.ucl.ac.be}} and
Pierre~M. Van Hove\footnote{E-mail address: {\tt pvh@fynu.ucl.ac.be}}

\vspace{0.5cm}

{\em Institut de Physique Nucl\'eaire}\\
{\em Universit\'e catholique de Louvain}\\
{\em B-1348 Louvain-la-Neuve, Belgium}\\

\vspace{2cm}

\begin{abstract}

\noindent An anomaly in the time behaviour of the signals observed by the
KARMEN Col\-la\-bo\-ra\-tion may be interpreted as 
the possible decay signature of a 33.9 MeV/c$^2$ mainly sterile 
neutrino. This note discusses the parameter
space still open for the mixing of this hypothetical particle with the
neutrinos of known leptonic flavour, considering the experimental results
which became available recently, as well as those to be expected
from forthcoming measurements. It is concluded that if no positive
signature is observed, the envisaged laboratory experiments are
not expected to close enterily the parameter space of mixing amplitudes.
However, a proper reassessment of the ALEPH upper bound on the 
$\nu_\tau$ neutrino mass including the possibility of $\tau$ flavour
mixing, would certainly help in reducing the parameter space
left open.

\end{abstract}

\end{centering} 

\vspace{4cm}

\noindent UCL-IPN-96-P04\\
August 1996

\clearpage

\end{titlepage}

\setcounter{footnote}{0}

\noindent{\bf 1. Introduction.} An excess of events observed in the
KARMEN detector\cite{KARMEN} about 3.6 $\mu$s after the beam-on-target,
was tentatively interpreted as the decay signature 
of a 33.9 MeV/c$^2$ neutral particle emitted in pion decay\cite{KARMEN}.
The genuine physical reality of these events seems to be
further confirmed by still more such events from KARMEN\cite{Drexlin}.

Ref.\cite{Sarkar} addressed specifically the tentative
identification\cite{KARMEN} of the observed signal with a mainly isosinglet
(sterile) neutrino emitted in pion decay through its admixture into the
muonic flavour and decaying in flight principally into 
an electron-positron pair and an isodoublet neutrino\footnote{Another
interpretation within supersymmetric models has also been 
considered in Ref.\cite{Sarkar2}.}. The conclusion
reached in Ref.\cite{Sarkar} is that the observed decay rate
implies the following constraint on the mixing amplitudes
$U_{\alpha X}$ $(\alpha=e,\mu,\tau)$ of the 33.9 MeV/c$^2$ mass eigenstate $X$
with the known leptonic flavours, provided the quantities
$\Big|U_{\alpha X}\Big|^2$ $(\alpha=e,\mu,\tau$) are small
enough for this first order expanded approximation to a more
complete analysis to be justified,
\begin{equation}
0.0285\,K\,\Big|U_{\mu X}\Big|^2\,
\Bigg[\,920\,\Big|U_{eX}\Big|^2+210\,\Big|U_{\mu X}\Big|^2+
210\Big|U_{\tau X}\Big|^2\,\Bigg]\,=\,3\cdot 10^{-11}\ {\rm s}^{-1}\ \ \ .
\label{eq:KARMEN}
\end{equation}
Here, the parameter $K$ takes the value $K=1$ (resp. $K=2$) 
for a Dirac (resp. Majorana) spinor describing the $X$ neutrino.

An isodoublet characterisation of the $X$ particle, and its
identification with the dominant component of the $\nu_\tau$ neutrino, were
rejected on the basis of cosmological and astrophysical arguments\cite{Sarkar}
as well as the ALEPH bound of 24 MeV/c$^2$ (95 \%C.L.) on the $\nu_\tau$
neutrino mass\cite{ALEPH}. It should be noted that the validity
of that bound was questioned recently in the context of
neutrino mixing scenarios\cite{Guzzo} in the $\tau$ flavour sector.

Independently of this controversy, the aim of this note
is to investigate the impact of the negative outcome of recent
searches for the $X$ particle\cite{Daum1,Bilger,Bry}, as well as that
of a search soon to be pursued anew\cite{Daum2}, and to identify
possible laboratory experiments which could either confirm the
existence of this particle, or else, exclude part or all of the parameter
space implied by the KARMEN experiment on the basis of (\ref{eq:KARMEN}).

\vspace{10pt}

\noindent{\bf 2. The Sterile Neutrino Scenario Revisited.} 
Let us reconsider the sterile neutrino
scenario as a possible explanation for the KARMEN anomaly\cite{KARMEN,Sarkar},
wihout any assumptions as to the smallness of mixing amplitudes.
The neutrino flavour eigenstates $\nu_a$ are indexed by
$(a=\alpha,0)$ with $\alpha=e,\mu,\tau$. The $a=\alpha$ flavours
are the known electroweak $SU(2)_L$ isodoublet neutrinos, while the
$a=0$ flavour is a $SU(2)_L\times U(1)_Y$ singlet.
For simplicity, it is assumed that only one such isosinglet
state is involved in the KARMEN anomaly, namely that the anomaly
is dominated by only one such state.

Neutrino interaction eigenstates $\nu_a$ are mixtures of the neutrino mass 
eigenstates. These mass eigenstates are denote $\nu_I$, the index $I=(i,X)$
taking the values $i=1,2,3$ associated to the three known light
neutrinos, and a fourth value $I=X$ corresponding to the $X$ particle
of the KARMEN anomaly. All these states are related through a mixing
matrix $U_{aI}$ such that,
\begin{equation}
\nu_a=U_{aI}\,\nu_I\ \ \ ,\ \ \ 
\overline{\nu_a}=\overline{\nu_I}\,U^\dagger_{Ia}\ \ \ .
\end{equation}
If no other states are involved in such a mixing
scenario, the matrix $U_{aI}$ is unitary. However, if for example
there were to exist additional singlet sterile neutrinos mixing
with the usual neutrino flavours, but at a level such that their
production in pion decay would be suppressed compared to that of the $X$
particle, the matrix $U_{aI}$ would then not have to be unitary, but
rather, be a sub-matrix of still a larger unitary mixing matrix.
However, note that in the circumstance that the matrix $U_{aI}$
is indeed unitary, one has the identity,
\begin{equation}
\sum_{I,J}\Big|\sum_{\alpha}U^\dagger_{J\alpha}U_{\alpha I}\Big|^2=3\ \ \ ,
\label{eq:SumU}
\end{equation}
in which the intermediate sum is over the isodoublet flavours
$\alpha=e,\mu,\tau$ rather than over all four neutrino flavours $a$.

In the following, we shall make the approximation to ignore the
masses of all ordinary neutrinos $\nu_i$ $(i=1,2,3)$
which are assumed to be sufficiently
small in comparison to all other mass scales involved, including in
particular the mass $m_X$ of the $X$ particle. It is then
straightforward to compute some relevant decay rates.

First, the pion decay branching ratios are simply given by,
in an obvious notation,
\begin{equation}
\frac{\Gamma(\pi^+\rightarrow \ell^+_\alpha X)}
{\Gamma(\pi^+\rightarrow \ell^+_\alpha\nu_i)}=
\frac{\Big|U_{\alpha X}\Big|^2}{\Big|U_{\alpha i}\Big|^2}\,
\frac{\lambda^{1/2}(m^2_\pi,m^2_\ell,m^2_X)}{m^2_\ell}\,
\frac{\Big[\,m^2_\pi(m^2_\ell+m^2_X)-(m^2_\ell-m^2_X)^2\,\Big]}
{(m^2_\pi-m^2_\ell)^2}\ \ \ ,
\label{eq:Br}
\end{equation}
with the usual kinematical factor $\lambda(a,b,c)=a^2+b^2+c^2-2ab-2ac-2bc$.

Next, concerning $X$ decay modes, the electronic branch 
gives,
\begin{displaymath}
\Gamma(X\rightarrow \nu_i e^+ e^-)=
\Bigg\{\,\frac{1}{8}\Big|\sum_{\alpha}U^\dagger_{i\alpha}U_{\alpha X}\Big|^2
(1+\rho^2)+\Big|U^\dagger_{ie}U_{eX}\Big|^2-
\end{displaymath}
\begin{equation}
-\frac{1}{2}\Re{\rm e}
\Big[\Big(\sum_\alpha U^\dagger_{i\alpha}U_{\alpha X}\Big)
{\Big(U^\dagger_{ie}U_{eX}\Big)}^*\Big](1+\rho)\,\Bigg\}
\,K\,\Gamma_X\ \ \ ,
\label{eq:elec}
\end{equation}
where,
\begin{equation}
\Gamma_X=\frac{G^2_F m^5_X}{192 \pi^3}\ \ \ ,\ \ \
\rho=1-4\sin^2\theta_W\ \ \ ,
\end{equation}
$G_F$ being Fermi's coupling constant, and $\theta_W$ the electroweak
gauge mixing angle such that\cite{PDG} $\sin^2\theta_W=0.2319$.

Similarly, for the neutrino branch, one has,
\begin{equation}
\Gamma(X\rightarrow \nu_i\nu_j\overline{\nu_k}; i\ne j)=
\frac{1}{4}\,\Big|\sum_\alpha U^\dagger_{i\alpha}U_{\alpha X}\Big|^2\,
\Big|\sum_\beta U^\dagger_{j\beta}U_{\beta k}\Big|^2\,K\,\Gamma_X\ \ \ ,
\label{eq:neu1}
\end{equation}
\begin{equation}
\Gamma(X\rightarrow \nu_i\nu_j\overline{\nu_k}; i=j)=
\,\frac{1}{2}\,\Big|\sum_\alpha U^\dagger_{i\alpha}U_{\alpha X}\Big|^2\,
\Big|\sum_\beta U^\dagger_{i\beta}U_{\beta k}\Big|^2\,K\,\Gamma_X\ \ \ .
\label{eq:neu2}
\end{equation}

Finally, the radiative branch is given by\cite{Moh},
\begin{equation}
\Gamma(X\rightarrow \nu_i\gamma)\simeq\frac{27\alpha}{8\pi}\,
\Big|\sum_\alpha U^\dagger_{i\alpha}U_{\alpha X}\Big|^2\,K\,\Gamma_X\ \ \ .
\label{eq:rad}
\end{equation}

For later reference, the $Z_0$ neutrino decay rates are also useful,
\begin{equation}
\Gamma(Z_0\rightarrow \nu_I\overline{\nu_J})=
\frac{G_F M^3_Z}{12\pi\sqrt{2}}
\Big|\sum_\alpha U^\dagger_{J\alpha}U_{\alpha I}\Big|^2
\frac{\lambda^{1/2}(M^2_Z,m^2_I,m^2_J)}{M^2_Z}
\Big[\,1-\frac{m^2_I+m^2_J}{2M^2_Z}-\frac{(m^2_I-m^2_J)^2}{4M^4_Z}\,\Big]\ \ ,
\label{eq:Z0}
\end{equation}
which, in the limit that the neutrino masses are neglected compared
to the $Z_0$ mass $M_Z$, reduces to,
\begin{equation}
\Gamma(Z_0\rightarrow \nu_I\overline{\nu_J})=
\frac{G_F M^3_Z}{12\pi\sqrt{2}}\,\Big|\sum_\alpha U^\dagger_{J\alpha}
U_{\alpha I}\Big|^2\ \ \ .
\end{equation}
Note that in this limit and when the matrix $U_{aI}$ is assumed
unitary, the total invisible $Z_0$ decay width 
$\sum_{I,J}\Gamma(Z_0\rightarrow \nu_I\overline{\nu_J})$
reduces to the Standard Model result for three generations of
electroweak neutrino isodoublets, as follows from (\ref{eq:SumU}).

The time characteristics of the KARMEN anomaly correspond\cite{KARMEN} to a mass
$M_X$ of
\begin{equation}
M_X=33.905\ {\rm MeV}/c^2\ \ \ ,
\end{equation}
which is indeed very close to the $\pi^\pm-\mu^\pm$ mass difference
of 33.91157 MeV/c$^2$. Cor\-res\-pondingly, the phase space factor appearing
in the pion decay branching ratio (\ref{eq:Br}) is then given by,
\begin{equation}
Br=\frac{\Gamma(\pi^+\rightarrow\mu^+ X)}
{\sum_{I}\Gamma(\pi^+\rightarrow\mu^+\nu_I)}=
\frac{0.0293\,\Big|U_{\mu X}\Big|^2}{\Big|U_{\mu 1}\Big|^2
+\Big|U_{\mu 2}\Big|^2+\Big|U_{\mu 3}\Big|^2+0.0293\Big|U_{\mu X}\Big|^2}
\ \ \ ,
\end{equation}
while the quantity $\Gamma_X$ takes the value,
\begin{equation}
\Gamma_X=1556\ {\rm s}^{-1}\ \ \ .
\end{equation}

The quantity determined for the KARMEN anomaly is\cite{KARMEN,Sarkar},
\begin{equation}
Br\,\Gamma_{\rm vis}\simeq 3 \cdot 10^{-11}\ {\rm s}^{-1}\ \ \ ,
\label{eq:BrGammavis}
\end{equation}
where the visible decay width of the $X$ particle is defined by,
\begin{equation}
\Gamma_{\rm vis}=\sum_{i=1,2,3}\Gamma(X\rightarrow\nu_i e^- e^+)\,+\,
\sum_{i=1,2,3}\Gamma(X\rightarrow\nu_i\gamma)\ \ \ .
\label{eq:Gammavis}
\end{equation}

Recent experimental results\cite{Daum1,Bilger,Bry} imply upper bounds on the
branching ratio $Br$, the most stringent of which is\cite{Daum1},
\begin{equation}
Br\,<\,2.6\cdot 10^{-8}\ (95\%\ {\rm C.L.})\ \ \ .
\label{eq:bound}
\end{equation}
A recent proposal\cite{Daum2} aims at improving this limit
down to the level of $10^{-10}$.
A further experimental constraint in the electronic
sector is the upper bound
on $\Big|U_{eX}\Big|^2$ from Ref.\cite{Rosier},
\begin{equation}
\Big|U_{eX}\Big|^2\,\stackrel{<}\sim\,10^{-6}\ \ \ .
\label{eq:boundelec}
\end{equation}

\vspace{10pt}

\noindent{\bf Minimal Mixing.} A minimal mixing hypothesis,
involving a single sterile neutrino and
compatible with the emission of the $X$ particle jointly with a muon,
would imply a unitary mixing matrix of the form,
\begin{equation}
U_{aI}=\left(\begin{array}{c c c c}
	1 & 0 & 0 & 0 \\
	0 & e^{i\alpha}\cos\theta & 0 & e^{i\beta}\sin\theta \\
	0 & 0 & 1 & 0 \\
	0 & -e^{i\gamma} \sin\theta & 0 & e^{i(\gamma-\alpha+\beta)}\cos\theta
	\end{array}\right)\ \ \ ,
\label{eq:singlemixing}
\end{equation}
where $\theta$ is a mixing angle, and $\alpha$, $\beta$ and $\gamma$
arbitrary phase factors. In particular, note that this ansatz
does not lead to mixing of the $X$ particle 
with the electron or tau flavours.

The KARMEN result (\ref{eq:BrGammavis}) then implies the values
for the flavour mixing angle,
\begin{equation}
\begin{array}{r l}
{\rm Dirac\ case}\ K=1:&\ \ \ \sin^2\theta=2.2\cdot 10^{-6}\ \ \ ,\ \ \ 
\sin 2\theta=3.0\cdot 10^{-3}\ \ \ ,\\ \\
{\rm Majorana\ case}\ K=2:&\ \ \ \sin^2\theta=1.6\cdot 10^{-6}\ \ \ ,\ \ \ 
\sin 2\theta=2.5\cdot 10^{-3}\ \ \ ,\ \ \ 
\end{array}
\end{equation}
which in turn leads to the branching ratios,
\begin{equation}
\begin{array}{r l}
{\rm Dirac\ case}\ K=1:&\ \ \ Br=6.5\cdot 10^{-8}\ \ \ ,\\ \\
{\rm Majorana\ case}\ K=2:&\ \ \ Br=4.6\cdot 10^{-8}\ \ \ .
\end{array}
\end{equation}

Hence, this result being larger than the recent upper bound
(\ref{eq:bound}) by a factor 2.5
or 1.8 for $K=1$ or $K=2$, respectively,
the minimal mixing scenario is already excluded 
on purely laboratory experimental grounds.

\vspace{10pt}

\noindent{\bf The Isodoublet Diagonal Mixing Approximation.}
The minimal mixing hypothesis shows that if the sterile neutrino
interpretation of the KARMEN anomaly were to be viable, one
would have to expect rather small mixing of the $X$ particle
with the known leptonic flavours, but still large enough to
account for the KARMEN constraint (\ref{eq:BrGammavis}). 
This suggests the following approximation, namely that the admixture of
the sterile neutrino flavour $a=0$ into the known massive neutrino states
is dominant compared to the mixing of the usual flavours 
$\alpha=e,\mu,\tau$ into the same states. Correspondingly,
such an isodoublet diagonal mixing approximation amounts to the ansatz,
\begin{equation}
U_{aI}=\left(\begin{array}{c c c c}
		\sqrt{1-x_e} & 0 & 0 & U_{eX} \\
		0 & \sqrt{1-x_\mu} & 0 & U_{\mu X} \\
		0 & 0 & \sqrt{1-x_\tau} & U_{\tau X} \\
		-U^*_{eX} & -U^*_{\mu X} & -U^*_{\tau X} &
	\sqrt{1-x_e-x_\mu-x_\tau} 
	      \end{array}\right)\ \ \ ,
\label{eq:ida}
\end{equation}
where
\begin{equation}
x_e=\Big|U_{eX}\Big|^2\ \ \ ,\ \ \ 
x_\mu=\Big|U_{\mu X}\Big|^2\ \ \ ,\ \ \ 
x_\tau=\Big|U_{\tau X}\Big|^2\ \ \ .
\end{equation}
Even though such a mixing matrix is not unitary, leptonic flavour
unitarity is violated only in the off-diagonal entries of the matrices
$U U^\dagger$ and $U^\dagger U$, to order $U_{\alpha X}U^*_{\beta X}$
$(\alpha,\beta=e,\mu,\tau\,;\alpha\ne \beta)$ for the $(3\times 3)$
submatrices of the first three lines and columns of these matrices,
and to order $U_{\alpha X}\Big|U_{\beta X}\Big|^2$ 
$(\alpha,\beta=e,\mu,\tau\,;\alpha\ne \beta)$ for the last line and column of
these matrices. The diagonal entries of $U U^\dagger$ and $U^\dagger U$
are all equal to unity. In particular, these properties of
(\ref{eq:ida}) imply that the identity (\ref{eq:SumU}) is violated
by terms of order $x_\alpha x_\beta$ 
$(\alpha,\beta=e,\mu,\tau\,;\alpha\ne \beta)$ only.

As a matter of fact, the choice in (\ref{eq:ida}) is the closest
one may come to a unitary mixing matrix given the isodoublet diagonal
mixing approximation. In particular for $U_{eX}=0$ and $U_{\mu X}=0$,
the ansatz (\ref{eq:ida}) indeed defines a unitary matrix mixing the
$a=\tau$ and $a=0$ neutrino flavours only. Since the present
experimental upper bounds on $\Big|U_{eX}\Big|^2$ and $Br$ imply values
for $x_e$ and $x_\mu$ less than $10^{-6}$, the choice in (\ref{eq:ida})
is thus in fact rather close to being unitary.

This is to be contrasted with the weak mixing approximation implicitly
assumed in Ref.\cite{Sarkar}, corresponding in fact to the matrix
(\ref{eq:ida}) linearised to first order in $U_{\alpha X}$ 
$(\alpha=e,\mu,\tau)$, namely,
\begin{equation}
U_{aI}\simeq\left(\begin{array}{c c c c}
		1 & 0 & 0 & U_{eX} \\
		0 & 1 & 0 & U_{\mu X} \\
		0 & 0 & 1 & U_{\tau X} \\
		-U^*_{eX} & -U^*_{\mu X} & -U^*_{\tau X} & 1
	      \end{array}\right)\ \ \ .
\label{eq:singlecolumn}
\end{equation}
Clearly, since the sum of the modulus squared entries for each line
and column is already larger than unity, such a choice is not unitary,
nor can it be a submatrix of a still larger unitary matrix.
Therefore, the weak mixing approximation (\ref{eq:singlecolumn})
may be physically meaningfull only when all three mixing amplitudes
$U_{\alpha X}$ $(\alpha=e,\mu,\tau)$ are sufficiently 
small\footnote{As will appear later on, this is indeed the case for the
parameters $x_e$ and $x_\mu$, but not necessarily for the parameter
$x_\tau$, which on account of the KARMEN constraint (\ref{eq:BrGammavis}),
approaches the value unity the smaller $x_\mu$ is contrained to be.}.
In particular, it is only under this assumption that the constraint
(\ref{eq:KARMEN}) does apply. In contradistinction to (\ref{eq:ida}),
the closer $x_\tau$ is to unity, the larger is the violation of
leptonic unitarity in (\ref{eq:singlecolumn}),
even when $U_{eX}=0$ and $U_{\mu X}=0$. Compared to (\ref{eq:ida}),
(\ref{eq:singlecolumn}) violates leptonic unitarity
in the products $UU^\dagger$ and $U^\dagger U$ to order
$U_{\alpha X}U^*_{\beta X}$ $(\alpha,\beta=e,\mu,\tau)$, and the
identity (\ref{eq:SumU}) as well to order $x_\alpha$ $(\alpha=e,\mu,\tau)$.

Given the isodoublet diagonal mixing ansatz (\ref{eq:ida}),
the KARMEN result (\ref{eq:BrGammavis}) reduces to the constraint,
\begin{equation}
K\,\frac{0.0293\,x_\mu}{(1-x_\mu)+0.0293 x_\mu}\,
\Bigg[\,929x_e(1-x_e)+208x_\mu(1-x_\mu)+
208x_\tau(1-x_\tau)\,\Bigg]=3\cdot 10^{-11}\ \ \ .
\label{eq:KARMEN2}
\end{equation}
Note that this relation does indeed reduce\footnote{The numerics in this
expression are somewhat different from those in (\ref{eq:KARMEN}),
presumably due to a value for $M_X$ slightly different from the
unquoted one used in Ref.\cite{Sarkar}.} to (\ref{eq:KARMEN})
provided all three quantities $x_\alpha$ $(\alpha=e,\mu,\tau)$ are
sufficiently small.

On the other hand, the recent experimental upper bound (\ref{eq:bound})
implies the limit,
\begin{equation}
x_\mu\,<\,8.9\cdot 10^{-7}\ (95\%\ {\rm C.L.})\ \ \ .
\label{eq:bound2}
\end{equation}

Fig.~1 displays the contraint (\ref{eq:KARMEN2})
for a $X$ particle of Dirac or Majorana character. 
The curves represent the pairs of values for
$({\rm log}_{10}x_\mu,{\rm log}_{10}x_\tau)$ 
related through (\ref{eq:KARMEN2}) for two extreme values of
$x_e$, namely $x_e=0$ for the dashed curve and 
$x_e=10^{-6}$ for the dotted-dashed curve, corresponding to the
range (\ref{eq:boundelec}) presently allowed
by laboratory experiments\cite{Rosier}. Note that for 
$x_\mu\le 3\cdot 10^{-8}$, these two curves become
essentially undistinguishible. 

The vertical lines in Fig.~1 also indicate the 95\% C.L. upper bound 
(\ref{eq:bound2}) of 
${\rm log}_{10}x_\mu=-6.05$ inferred from Ref.\cite{Daum1},
as well as the limit of $x_\mu< 3.4\cdot 10^{-9}$,
corresponding to ${\rm log}_{10}x_\mu=-8.47$,
to be reached by a new experiment\cite{Daum2} which aims at constraining
the branching ratio $Br$ down to a level of $10^{-10}$ or better.
The associated values for $x_\tau$ are also indicated as dashed
or dotted-dashed horizontal lines, corresponding to
$x_e=0$ or $x_e=10^{-6}$, respectively. 
Note that in the $K=2$ Majorana case, the present upper bound
of $x_\mu< 8.9\cdot 10^{-7}$ implies that values
for $x_e$ as large as $10^{-6}$ are already excluded
on the basis of the KARMEN constraint (\ref{eq:KARMEN2}).
For the soon to be expected\cite{Daum2} upper bound of 
$x_\mu=3.4\cdot 10^{-9}$,
the value for $x_\tau$ is independent of $x_e< 10^{-6}$.
Note that these values for $x_\tau$ are lower bounds
on that parameter, implied by the KARMEN constraint (\ref{eq:KARMEN2}).
The canonical upper bound on $x_\tau$ is
of course the value unity, which in turn leads to the lower bound
on $x_\mu$ of $2.0\cdot 10^{-11}$ in the $K=1$ Dirac
case, or $9.8\cdot 10^{-12}$ in the $K=2$ Majorana case.
Note that the existence of such non vanishing lower bounds on the
mixing parameter $x_\tau$ also explains {\sl at posteriori\/}
why the minimal mixing scenario of the previous section---in which
$U_{\tau X}=0$---is inconsistent with the presently most stringent 
upper bound (\ref{eq:bound}) on $Br$.

{}From Fig.~1, one may thus conclude that, assuming the sterile
neutrino in\-ter\-pre\-ta\-tion\cite{KARMEN,Sarkar} of the KARMEN anomaly
to be realised,
\begin{enumerate}
\item[i)] given the forthcoming measurements of Ref.\cite{Daum2},
there is no interest, from the present point of view, 
in improving the upper limit on $x_e\stackrel{<}\sim 10^{-6}$ 
of Ref.\cite{Rosier};
\item[ii)] at present, the $x_\mu$ mixing parameter
must be less than $8.9\cdot 10^{-7}$ and larger than
$2.0\cdot 10^{-11}$ in the Dirac case or $9.8\cdot 10^{-12}$
in the Majorana case;
\item[ii)] unless a positive signal is found, even the forthcoming
measurements\cite{Daum2} will not be able to exclude
the sterile neutrino scenario with the mixing amplitude squared
$x_\mu$ in the range from $3.4\cdot 10^{-9}$ to 
$2.0\cdot 10^{-11}$ in the $K=1$ Dirac case
or $9.8\cdot 10^{-12}$ in the $K=2$ Majorana case,
or correspondingly, values for $x_\tau$
between unity and $1.45\cdot 10^{-3}$ in the Dirac case, or
$7.24\cdot 10^{-4}$ in the Majorana case. The latter conclusions
are independent of the value for the mixing parameter $x_e< 10^{-6}$.
\end{enumerate}
In this scenario, the $X$ particle of
33.9 MeV/c$^2$ would be emitted in pion decay due to its admixture into
the muonic flavour, and would decay inside the detector essentially due to its
admixture into the $\tau$ flavour via the neutral current coupling
to an electron-positron pair. At present, it seems difficult to
push further such limits on this interpretation of
the KARMEN anomaly through laboratory experiments.

\clearpage

\vspace{10pt}

\noindent{\bf The $Z_0$ Invisible Decay Width and $\tau$ Decays.} 
In order to reach more definite conclusions with regards
to the sterile neutrino scenario of Ref.\cite{Sarkar},
it seems worthwhile to explore additional experimental laboratory constraints.

A first possibility which may come to mind is that of the
invisible $Z_0$ decay width. In fact, since the isodoublet diagonal
mixing ansatz (\ref{eq:ida}) breaks leptonic unitarity, 
and consequently violates the
identity (\ref{eq:SumU}) to order $x_\alpha x_\beta$ 
$(\alpha,\beta=e,\mu,\tau\,;\alpha\ne \beta)$, a deviation from the
Standard Model value for the $Z_0$ invisible width should involve
a combination of these products of mixing parameters. Indeed, 
using the experimental values\cite{PDG},
\begin{equation}
\Gamma_{\rm inv}(Z_0)=498.3\pm 4.2\ {\rm MeV}\ \ \ ,\ \ \ 
M_Z=91.187\pm 0.007\ {\rm GeV}/c^2\ \ \ ,
\end{equation}
the result (\ref{eq:Z0}) implies the constraint,
\begin{equation}
x_e x_\mu+x_e x_\tau+x_\mu x_\tau= 0.0017\pm 0.013\ \ \ 
\label{eq:bound3}.
\end{equation}
Quite clearly, given the present or even 
the forthcoming experimental upper bounds
on the parameters $x_e$ and $x_\mu$ of $10^{-6}$ or less, even if
the $\tau$ mixing parameter $x_\tau$ were to reach its maximal value of unity,
this constraint is satisfied to many orders of magnitude.

In contradistinction, since the weak mixing approximation 
(\ref{eq:singlecolumn}) violates the identity (\ref{eq:SumU})
to order $x_\alpha$ $(\alpha=e,\mu,\tau)$, the $Z_0$ insivible decay
width constraint reads in that case,
\begin{equation}
x_e+x_\mu+x_\tau=0.0017\pm 0.013\ \ \ .
\label{eq:bound4}
\end{equation}
Taken at face value, this condition would imply the additional
upper bound on $x_\tau$ of $x_\tau < 0.013$, or ${\rm log}_{10}x_\tau=-1.89$,
thereby narrowing down further the window which is still
open in the $(x_\mu,x_\tau)$ parameter space from the discussion
of the previous section, even when the forthcoming limit of
Ref.\cite{Daum2} is accounted for. However, for the reasons
explained previously, the weak mixing ansatz (\ref{eq:singlecolumn})
is not physically acceptable, and the constraint (\ref{eq:bound4})
can therefore not be included in any analysis.

Since the parameter $x_\tau$ is not
constrained in the analysis of the previous section,
another possibility which may be contemplated is that of $\tau$ decays.
However, a simple consideration of the muonic and electronic
branching ratios\cite{PDG} within the isodoublet diagonal
mixing approximation implies that the ratio of these two
branching ratios, when properly accounting for leptonic mass
contributions, should reduce to unity,
whereas the experimental data\cite{PDG} for the likewise mass corrected
ratio lead to the result $1.009\pm 0.017$. Hence, no additional
constrained is to be derived from these leptonic branching ratios.
Incidentally, within the weak mixing approximation, the same
ratio provides for the constraint
$(1+x_\mu/1+x_e)=1.009\pm 0.017$,
which clearly, is easily accomodated given the present limits
of $x_\mu< 8.9\cdot 10^{-7}$ (95\% C.L.) and 
$x_e\stackrel{<}\sim 10^{-6}$,
independently of the fact that the weak mixing ansatz 
(\ref{eq:singlecolumn}) is physically unsatisfactory.

It is conceivable nevertheless that $\tau$ decays may provide
for a constraint on the mixing parameter $x_\tau$. Indeed,
in Ref.\cite{Guzzo}, it is argued that when considering the
possibility of mixing in the $\tau$ leptonic sector, the
ALEPH upper bound\cite{ALEPH} of 24 MeV/c$^2$ on 
the mass of the $\nu_\tau$ neutrino
has to be reconsidered. However, within the isodouble diagonal mixing
ansatz (\ref{eq:ida}) and when the parameter $x_\tau$ is very close
to unity, a situation still allowed on basis of the analysis
of the previous section, it is mostly the $X$ particle at 33.9 MeV/c$^2$
which couples to the $\tau\rightarrow 5\pi(\pi^0)\nu$
decay mode used by ALEPH, rather than the $\nu_3$ mass eigenstate.
{}From the point of view of the kinematical analysis leading
to the ALEPH mass limit, whether it is the $X$ particle or the
$\nu_3$ neutrino does not make a difference, so that the ALEPH
result should in fact lead to an upper bound on $x_\tau$ somewhat
less that unity. However, the specific value for this upper bound
requires a detailed analysis of the ALEPH hadronic spectra
which would have to properly account for a mixing parameter
$x_\tau$ neither very close to unity, nor very small, along
lines similar to those advocated in Ref.\cite{Guzzo}.
On the basis of the statistics accumulated\cite{ALEPH},
it seems reasonable to expect that such a reassessment---which is
thus called for---of the ALEPH result, would be
able to reach an upper bound on the order of $10^{-2}$ or better
on $x_\tau$, namely ${\rm log}_{10}x_\tau=-2$, thereby reducing further
the window still open in the $(x_\mu,x_\tau)$ parameter space.

As a matter of fact, if such a detailed reanalysis of the ALEPH data
were to be completed, and if the sensitivity of the forthcoming
experiment\cite{Daum2} aiming for an upper limit of $10^{-10}$
on $Br$ were to be improved still to some extent, it may even become
possible to confirm, or either refute definitely the interpretation
of the KARMEN anomaly as being due to a massive isosinglet neutrino
mixing with the ordinary isodoublet neutrino flavours.

\vspace{10pt}

\noindent{\bf Acknowledgements.} 
M.~Daum, J.-M.~G\'erard and S.~Sarkar are gratefully acknow\-ledged 
for useful remarks and discussions on matters pertaining to the present note.

\pagebreak

\noindent{\bf Figure Caption}

\begin{itemize}
\item[Fig.~1:] The parameter space for muon and tau leptonic flavour
mixing with a sterile neutrino in the case of a Dirac (bottom) and of
a Majorana (top) $X$ particle. The dashed and dotted-dashed curves
represent the KARMEN constraint (\ref{eq:KARMEN2}) for
$\Big|U_{eX}\Big|^2=0$ and $\Big|U_{eX}\Big|^2=10^{-6}$, respectively. 
The vertical and horizontal lines are discussed in the text; 
they represent different upper or lower bounds on the mixing parameters,
with the shaded regions being the excluded ones.
\end{itemize}


\clearpage

\begin{thebibliography}{99}

\bibitem{KARMEN} B.~Armbruster {\it et al.\/}, {\sl Phys. Lett.\/}
{\bf B348} (1995) 19.

\bibitem{Drexlin} B.~Seligmann, in Proc. Int. Europhysics Conf. on
High Energy Physics, Brussels, 27 July - 2 August 1995,
eds.~J.~Lemonne, C.~Vander Velde and F.~Verbeure (World Scientific, Singapore,
1996), pp. 526-527;\\
G.~Drexlin, private communication.

\bibitem{Sarkar} V.~Barger, R.J.N.~Phillips and S.~Sarkar,
{\sl Phys. Lett.\/} {\bf B352} (1995) 365; (E) {\sl ibid}
{\bf B356} (1995) 671;\\
S.~Sarkar, communication at the Int. Europhysics Conf. on High Energy Physics,
Brussels, 27 July  - 2 August 1995.

\bibitem{Sarkar2} D.~Choudhury and S.~Sarkar,
{\sl Phys. Lett.} {\bf B374} (1996) 87.

\bibitem{ALEPH} ALEPH Collaboration, {\sl Phys. Lett.\/}
{\bf B349} (1995) 585.

\bibitem{Guzzo} M.M.~Guzzo, O.L.G.~Peres, V.~Pleitez and
R.~Zukanovich Funchal, {\sl Phys. Rev.\/} {\bf D53} (1996) 2851.

\bibitem{Daum1} M.~Daum {\it et al.\/}, {\sl Phys. Lett.} {\bf B361}
(1995) 179.

\bibitem{Bilger} R.~Bilger {\it et al.\/}, {\sl Phys. Lett.\/}
{\bf B363} (1995) 41.

\bibitem{Bry} D.A.~Bryman and T.~Numao, {\sl Phys. Rev.\/}
{\bf D53} (1996) 558.

\bibitem{Daum2} M.~Daum {\it et al.\/}, {\sl Search for a Neutral
Particle of Mass 33.9 MeV in Pion Decay\/}, Proposal for
an experiment at the Paul Scherrer Institute, June 1996.

\bibitem{PDG} Review of Particle Properties:
R.M.~Barnett {\it et al.\/} (Particle Data Group) {\sl Phys. Rev.}
{\bf D54} (1996) 1.

\bibitem{Moh} For references to the original literature as well
as a discussion, see for example,\\
R.N.~Mohapatra and P.B.~Pal, 
{\sl Massive Neutrinos in Physics and
Astrophysics\/} (World Scientific, Singapore, 1991).

\bibitem{Rosier} N.~De Leener-Rosier {\it et al.\/}, {\sl Phys. Rev.}
{\bf D43} (1991) 3611.



\end{thebibliography}
\end{document}